\documentclass[symmetry,article,accept,moreauthors,pdftex]{Definitions/mdpi} 

\firstpage{1} 
\pubvolume{xx}
\issuenum{1}
\articlenumber{5}
\pubyear{2020}
\copyrightyear{2020}
\history{Received: date; Accepted: date; Published: date}

\usepackage{bm}

\newcommand{\beq}{\begin{equation}}
\newcommand{\eeq}{\end{equation}}
\newcommand{\beqa}{\begin{eqnarray}}
\newcommand{\eeqa}{\end{eqnarray}}
\newcommand{\pr}{^{\prime}}
\newcommand{\ket}[1]{\left|#1\right>}
\newcommand{\bra}[1]{\left<#1\right|}
\newcommand{\braket}[1]{\left<#1\right>}

\newcommand{\eq}[1]{(\ref{#1})}
\def\bbbone{{\mathchoice {\rm 1\mskip-4mu l} {\rm 1\mskip-4mu l} {\rm 1\mskip-4.5mu l} {\rm 1\mskip-5mu l}}}

\usepackage{xcolor}
\newcommand{\Al}[1]{#1}
\newcommand{\Ma}[1]{#1}

\Title{Non-Hermitian Chiral Magnetic Effect In Equilibrium}

\Author{M.N. Chernodub $^{1,2,*}$\orcidA{} and Alberto Cortijo $^{3,4,*}$\orcidB{} }

\AuthorNames{Maxim Chernodub and Alberto Cortijo}

\address{%
$^{1}$ \quad Institut Denis Poisson UMR 7013, Universit\'e de Tours, 37200 France; maxim.chernodub@idpoisson.fr\\
$^{2}$ \quad Pacific Quantum Center, Far Eastern Federal University, Sukhanova 8, Vladivostok, 690950, Russia\\
$^{3}$ \quad Instituto de Ciencia de Materiales de Madrid, CSIC, Cantoblanco, 28049 Madrid, Spain\\
$^{4}$\quad Departamento de Fisica de la Materia Condensada, Universidad Autonoma de Madrid, Madrid E-28049, Spain; alberto.cortijo@uam.es\\
${}^*$ These authors contributed equally to this work.
}

\abstract{
We analyze the Chiral Magnetic Effect for non-Hermitian fermionic systems using the bi-orthogonal formulation of quantum mechanics. In contrast to the Hermitian counterparts, we show that the Chiral Magnetic effect takes place in equilibrium when a non-Hermitian system is considered. The key observation is that for non-Hermitian charged systems, there is no strict charge conservation as understood in Hermitian systems, so the Bloch theorem preventing currents in the thermodynamic limit and in equilibrium does not apply.}


\begin{document}


\section{Introduction} 
\Ma{The Chiral Magnetic Effect (CME) is the generation of an electric current  $\bm J$ in presence of an external magnetic field $\bm B$~ \cite{FKW08}}:
\beq
\braket{\bm{J}}=\frac{e^2}{2\pi^2}\mu_5\bm{B}.\label{CMEdef}
\eeq
\Ma{The current~\eq{CMEdef} appears naturally in a particular set of physical systems characterized by a broken invariance under the spatial reflection $\mathcal P$. The broken $\mathcal P$--invariance may be realized, for example, in condensed matter systems with massless fermionic quasiparticles of a Weyl or Dirac type. Such fermions are characterized by different (left and right) chiralities which are often said to be ascribed to different ``two Weyl nodes''. If the number of fermions with different chiralities is not equal to each other, then the system is $\mathcal P$--broken. The chiral imbalance is convenient to characterize by a difference denoted by $\mu_5=\mu_L-\mu_R$ between the chemical potentials in the right ($\mu_R$) and left ($\mu_L$) Weyl nodes. The difference in the chemical potentials determines the magnitude of the CME current in Eq.~\eq{CMEdef}.}

\Al{The CME is a relevant transport phenomenon that has its roots in the physics of quantum anomalies. The theory is said to be anomalous if there exists a quantity which is conserved at the classical level and which fails to do so when going to the quantum realm. In particular, the CME stems from the axial anomaly which leads to non-conservation of the chiral current in Weyl systems described by the Weyl Hamiltonian of a massless particle with the wavefunction~$\psi_s$:}
\beq
H_s = s v_F \psi^{+}_s\bm{\sigma}\cdot\bm{k}\psi_s.\label{Weyl1}
\eeq
\Ma{Here the parameter $s=\pm1$ denotes the chirality of the particle propagating with the velocity~$v_F$ and momentum $\bm k$ and $\bm \sigma$ is the vector of the Pauli matrices. In relativistic systems, Lorentz invariance forces $v_F=c$, while in condensed matter systems, the velocity $v_F$ is not constrained to any particular value. It is often said that the parameter $s=\pm1$ labels two distinct ``Weyl points''.}

\Al{The Weyl systems~(\ref{Weyl1}) possess two quantities that are conserved at the level of classical equations of motion. These are the electric charge $Q$ and the axial charge $Q_5$ that are described, respectively, by electric four-current $J^\mu$ and the chiral current\footnote{\Ma{We reserve the notation $\bm j$ for another current to be defined below~\eq{conservedcurrent}.}} $j^\mu_5$: 
\beq
J^\mu \equiv (Q,{\bm J}) = {\bar \Psi} \gamma^\mu \Psi, 
\qquad
j^{\mu}_5 \equiv (q_5,{\bm j}_5) = {\bar \Psi} \gamma_5\gamma^{\mu} \Psi.
\label{eq:j:5}
\eeq
Mathematically, the conservation implies that the four-divergence of the both currents is identically zero, $\partial_\mu J^\mu = \partial j^\mu_5= 0$, provided the wave function $\Psi = (\psi_R, \psi_L)^T$ satisfies the classical equation of motion $H \Psi = 0$ where the Hamiltonian $H = {\mathrm{diag}}\,(H_R,H_L)$ incorporates both chiral modes~\eq{Weyl1}. We use the conventional nomenclature of $\gamma$ matrices in the chiral basis:
\beq
\gamma^0 = \left(\begin{matrix} 0 & \bbbone \\ \bbbone & 0 \end{matrix}\right),
\qquad
{\bm \gamma} = \left(\begin{matrix} 0 & {\bm \sigma} \\ - {\bm \sigma} & 0 \end{matrix}\right),
\qquad
\gamma^5 = \left(\begin{matrix} - \bbbone & 0 \\ 0 & \bbbone \end{matrix}\right).
\label{eq:gamma}
\eeq
In presence of conventional electromagnetic fields, the quantum fluctuations lead to nonconservation of the chiral current. Technically, the loss of conservation appears as a result of the so-called triangle diagrams of virtual fermions that lead to~\cite{A69,BJ69}:}
\beq
\partial_{\mu}j^{\mu}_5=\frac{1}{2\pi^2}\bm{E}\cdot\bm{B}.\label{chiralanom}
\eeq

\Al{The triangle diagrams that give rise to the non-conserved current (\ref{chiralanom}) are also responsible for the generation of the current in Eq.(\ref{CMEdef}). The form of the anomaly can be partially fixed by some algebraic constraints on an effective action of the theory that leads the right hand side of Eq.(\ref{chiralanom}). These constraints are imposed by the algebraic structure of the symmetry that is anomalously broken~\cite{Kharzeev:2013ffa}. Technically, the relation of the chiral (triangular) anomaly to the generation CME current~\eq{CMEdef} requires a rigorous derivation which takes into account the Wess-Zumino consistency conditions~\cite{L16}.} \Ma{In the presence of the chiral gauge fields $A^{5}_{\mu}$ that are coupled to the chiral current $j^{\mu}_5$, the anomalous effects become more subtle and the currents~\eq{eq:j:5} have to be modified consistently. In our work $A^5_\mu \equiv 0$ so that we will use the straightforward definition~\eq{eq:j:5} for vector and axial currents.}

\Ma{In general, two ways have been proposed to create an environment that is able to generate the anomalous current~\eq{CMEdef}. The first approach is to drive the system out of equilibrium in order to reach a stationary regime where $\mu_5\neq0$.  This regime may be achieved by applying simultaneously an electric field $\bm{E}$ parallel to $\bm{B}$ so that the chiral anomaly creates a charge imbalance via the chiral anomaly~\eq{chiralanom} and generates a nonzero chiral chemical potential $\mu_5$. Then the system generates a non-equilibrium electric current via the CME mechanism~\eq{CMEdef}. Notice that the chiral imbalance $\mu_5 \neq 0$ does not exist in an equilibrium regime as the population of the left-handed particles and the right-hand particles mix with each other due to interactions and then relaxes towards $\mu_5 = 0$ equilibrium. 
Therefore, the anomalous electric current~\eq{CMEdef} is zero in thermal equilibrium.}

\Ma{The second approach could consist in moving the position of the Weyl nodes in energies without carrying the system out of equilibrium}:
\beq
\braket{\bm{J}}=\frac{e^2}{4\pi^2}(\epsilon_{R}-\epsilon_{L})\bm{B}=0,\label{CMEwrong}
\eeq
\Ma{where the energies $\epsilon_{L,R}$ are the position of left and right Weyl points in energy. The equilibrium current~\eq{CMEwrong} is, however, vanishing  in Hermitian systems. Physically, the current is zero because the difference in energies of the right- and left-handed chiral fermions does not create the true chiral imbalance. Mathematically, the current~\eq{CMEwrong} does not exist because the difference in energies $(\epsilon_{R}-\epsilon_{L})$ is sensitive to the chirality of the fermion and, therefore, is nothing but the zeroth component of a chiral gauge field, $A^5_0$. Once the chiral gauge field appears, the definition of the physical electric current starts to differ from the naive covariant version~\eq{eq:j:5} by an addition of an extra term coming from the so-called Bardeen polynomials. This term cancels out this energy difference precisely, and physical (so called, ``consistent'') version of the current, vanishes in thermal equilibrium~\eq{CMEwrong}. For further details and technicalities we refer an interested reader to Ref.~\cite{L16}.
 }

\Ma{Despite in thermodynamic equilibrium the axial chemical potential is zero, $\mu_5 = 0$, the vector chemical $\mu = \mu_R + \mu_L$ for a generic fermionic system may still be nonzero. In the presence of the background magnetic field $\bm B$, the system generates -- via the same chiral anomaly -- the chiral current:
\beq
\braket{\bm{J}_5}=\frac{e^2}{2\pi^2}\mu\bm{B},\label{CSEdef}
\eeq
which is the direct analogue of the CME~\eq{CMEdef} but now the chiral sector. Equation~\eq{CSEdef} describes the Chiral Separation Effect (CSE) which will play a role in our derivations below along with the CME.}

\begin{figure}
\begin{tabular}{cc}
\includegraphics[scale=0.2]{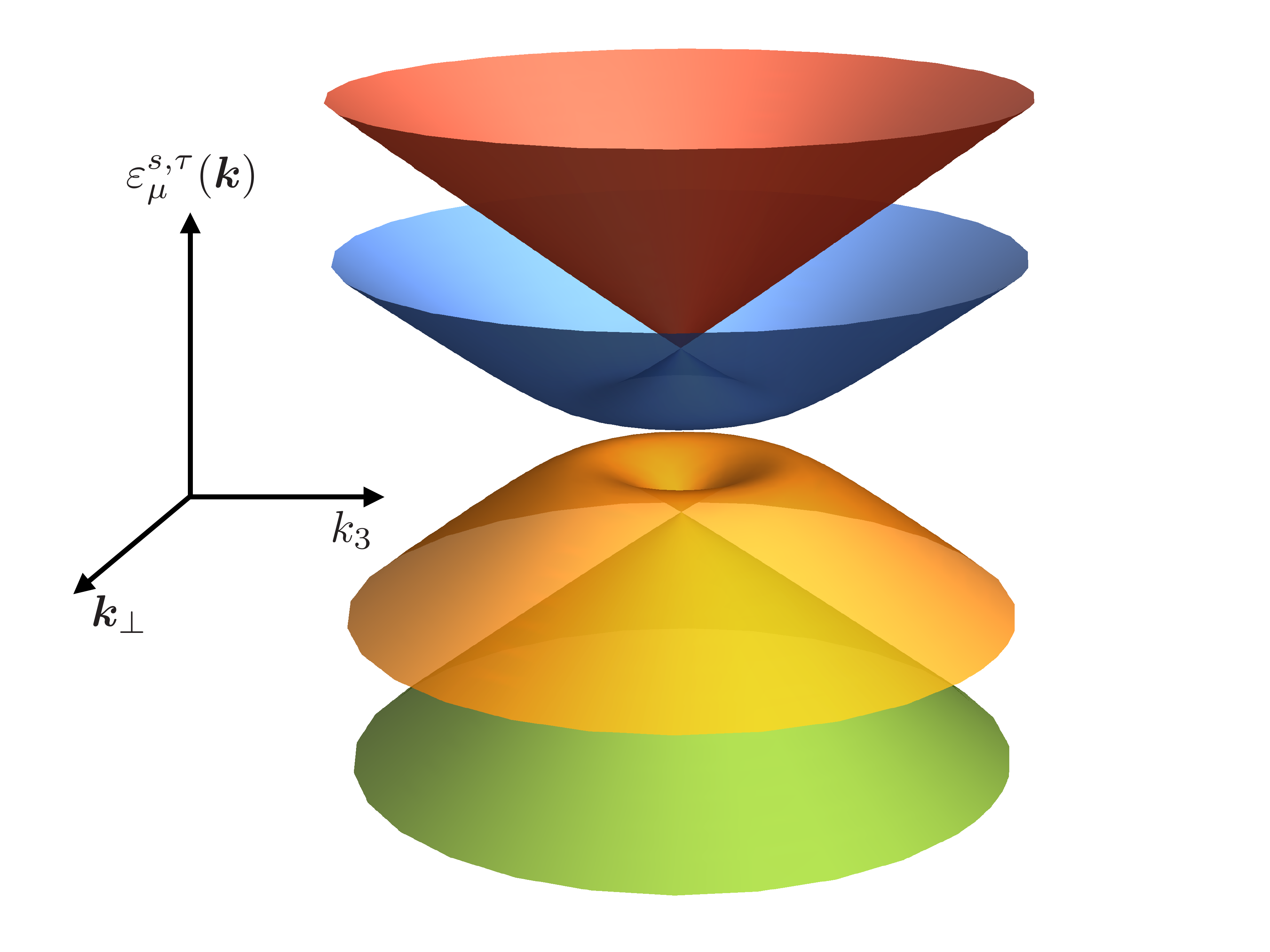} & \\[-50mm]
& \includegraphics[scale=0.2]{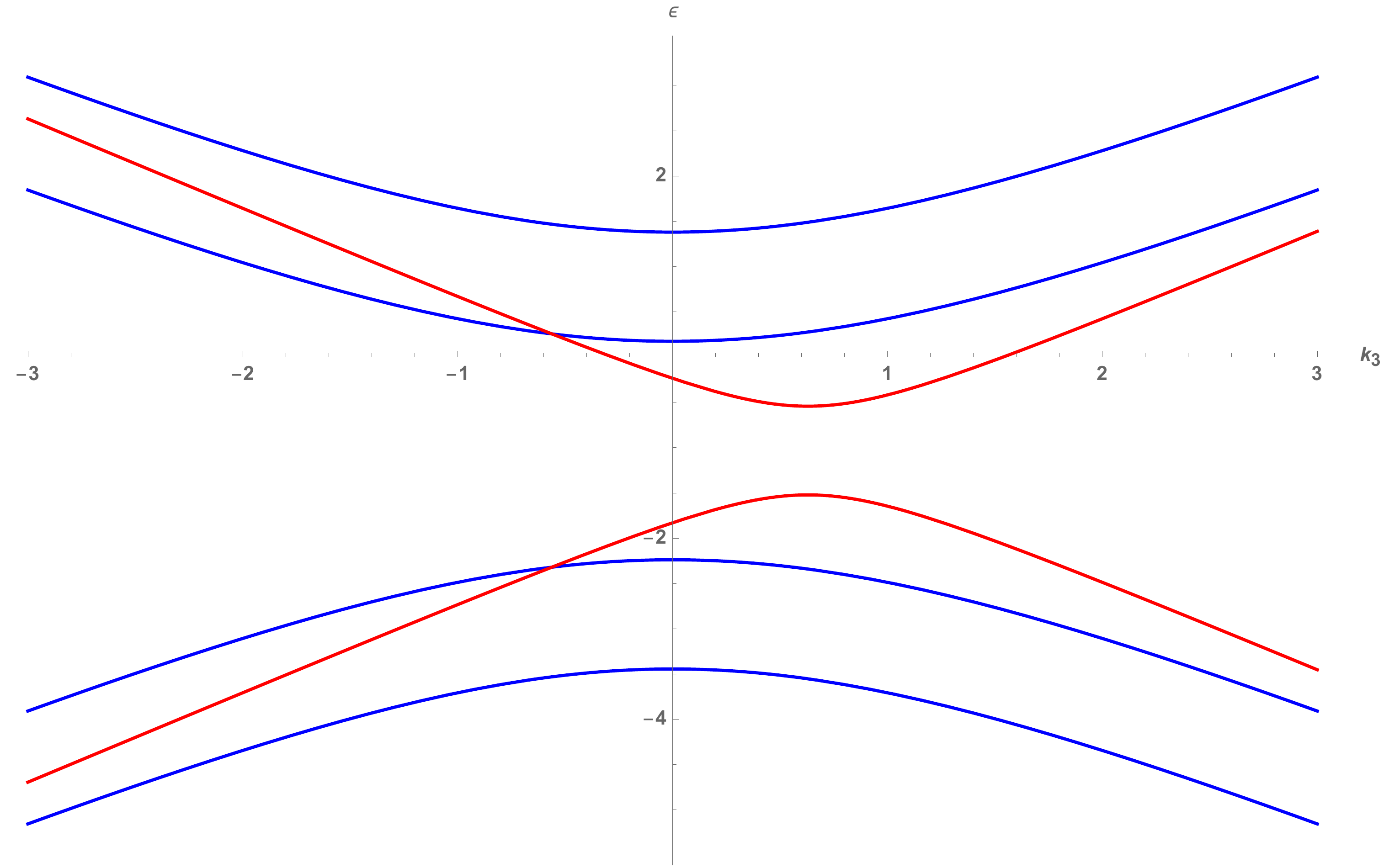} \\[10mm]
(a) & (b)
\end{tabular}
\caption{(color online) (a) Band structure of $\mathcal{H}$ at zero magnetic field but finite chemical potential. Contrary to Hermitian systems, the presence of chemical potentials might modify strongly the spectrum. (b) Landau level spectrum for the non-Hermitian model for finite chemical potential $\mu$. Finite values of $\mu$ shifts the LLL spectrum (red) not only upwards or downwards, but it is laterally shifted. The lateral shift makes the contribution from LLL to be non-zero for the CME.}
\label{fig:bands}
\end{figure}

While the non-equilibrium situation has been explored extensively in the literature leading, for instance,  to the celebrated negative quadratic magnetoresistivity in Weyl metals, the equilibrium scenario appears to be not possible, and to date there is consensus that the CME is not possible in equilibrium~\cite{MP15,Y15,Z16}. 

The statement of absence of CME in equilibrium can be seen as an extension of a no-go theorem given by Bloch, concerning the existence of equilibrium currents in solids in the thermodynamic limit\cite{B49}. This theorem has been extended to chiral matter in Ref.~\cite{Y15}, and refined in Ref.~\cite{Z16} (the absence of CME in equilibrium using the chiral kinetic formalism has been obtained in Ref.~\cite{MP15}). There are three elements usually associated to this theorem in chiral matter: The existence of Weyl nodes that always come by pairs\cite{NN181},\cite{NN281}, (local) gauge invariance and, of course, the assumption that the system is in the equilibrium state. As, we have mentioned, it is known how to break the second condition and drive the system out of equilibrium. Recently, it has been proposed that the first assumption of having pairs of Weyl nodes can be broken in Weyl superconductors, where an external magnetic field indices a gap in one of the Weyl nodes (and its particle-hole conjugate), leaving effectively a single Weyl node\cite{OBA17}. However, we stress here that the presence of Weyl nodes is not a strict requirement to the absence (or presence) of CME~\cite{Y15,Z16}.


Interestingly, non-Hermitian fermionic systems appear to be a 
\Ma{promising physical environment that can be realized in real experiments.} 
Nowadays, there is a surge of interest in non-Hermitian systems for many different reasons, ranging from very fundamental questions in the quantum (and statistical) theory of fields and the role of topology in non-Hermitian systems\cite{B05,C17,GAK18}, to applied science. Among them, specially interesting are the non-Hermitian systems that display a real spectrum, as the $\mathcal{P}*\mathcal{T}$-symmetric systems or the quasi-Hermitian systems. Although non-Hermitian, they display a unitary evolution, and it is possible to define a consistent thermodynamics for them\cite{GDS16}.


\section{The model} We will study is a non-Hermitian extension of the massive Dirac model in $(3+1)$ dimensions, where, together with the usual mass term $m$, an anti-Hermitian mass $m_5$ is introduced~\cite{AB15,ABM15,AMS17}:
\beq
H=\bm{\alpha}\cdot\bm{k}+m\hat{\beta}+m_5\hat{\beta}\gamma_5. \label{NHWeylHam}
\eeq
\Ma{Here we used the original Dirac notations ${\bm \alpha} \equiv \gamma^0 {\bm \gamma}$ and $\hat{\beta} \equiv \gamma^0$ where the Dirac gamma matrices are given in Eq.~\eq{eq:gamma}, and $\bm k$ describes the momentum of the particle.} The advantage of this model is that the two first terms of the right hand side of Eq.(\ref{NHWeylHam}) are Hermitian by themselves, so the only non-Hermitian (anti-Hermitian) term is $m_5\gamma_5$. \Al{When $m_5$ the model Eq.(\ref{NHWeylHam}) corresponds to the usual Dirac model for relativistic fermions. Also, it constitutes the low energy model for the bulk states of topological insulators, and when $m=0$ or $m=m(\bm{k})$ becomes a function with nodal points in momentum space, this model describes Weyl fermions \cite{AMV18}.}

\Al{To date, there is not known experimental realization of an electronic system with the non-Hermitian mass term $m_5\hat{\beta}\gamma_5$. } \Ma{However, we are not aware of any no-go theorem that would forbid this term to appear in open systems. Therefore, we consider the model~\eq{NHWeylHam} as a generic system which captures the essential properties of the non-Hermitian mass. Our aim is to show, conceptually, that the equilibrium Chiral Magnetic Effect is, in principle, possible in a generic non-Hermitian system.}

It is already stated in the literature that non-Hermitian hamiltonians are not gauge-invariant in general. This can be viewed as the fact that the Noether theorem relating continuous symmetries and conserved currents in field theories, does not hold in non-Hermitian systems\cite{ABM15,AMS17,M18}. For this reason, there is some arbitrariness when defining a coupling to electromagnetic fields in the Hamiltonian (\ref{NHWeylHam}). In the present work, we will interested to compare our results with the ones in Hermitian systems, so we will consider the coupling to electromagnetic fields to the model  (\ref{NHWeylHam}) with $m_5=0$, which is Hermitian (and the principle of local gauge invariance holds), and later we switch on the non-Hermitian term $m_5\hat{\beta}\gamma_5$.

Also, that we cannot apply the Noether theorem for Hermitian systems to non-Hermitian ones does not imply that conserved currents  exist for the latter.

\Al{In a conventional Hermitian quantum mechanics, the time dependence of any operator $\mathcal{O}$ can be determined in the Heisenberg picture:}
\beq
\mathcal{O}(t)=e^{iH^{+}t}\mathcal{O}e^{-iH t}.\label{Heis1}
\eeq
\Al{If we naively try to proceed further following the same steps for a NH system, we will get an unconventional expression for the time variation of the operator $\mathcal{O}(t)$\cite{GHK10,SZ13}:}
\beq
\dot{\mathcal{O}} \equiv \frac{ d\mathcal{O}(t)}{d t}=ie^{iH^{+}t}\left(H^+\mathcal{O}-\mathcal{O}H\right)e^{-iH t}.\label{TevolNH}
\eeq

For Hermitian systems, $H^+=H$ and we recognize the commutation with $H$ as the condition any operator must satisfy to be a  conserved quantity. For non-Hermitian systems, we immediately see that an operator is a conserved quantity if, instead of commuting with the Hamiltonian, it fulfills the quasi-Hermiticity condition: $H^+\mathcal{O}=\mathcal{O}H$ so $\dot{\mathcal{O}}=0$. In the case of the $U(1)$ charge symmetry, it is clear that the generator of this symmetry in Hermitian systems commutes with $H$ but does not satisfy the quasi-Hermiticity condition, so it is not a conserved quantity for non-Hermitian systems. As we will see in next paragraphs, it is possible to find operators that, not commuting with $H$, they satisfy the quasi-Hermiticity condition, thus defining conserved quantities.

Before computing the CME and CSE for the system at hands, it is convenient to understand what is the physical meaning of the conserved currents associated to the Hamiltonian (\ref{NHWeylHam}). The most convenient way is the following\cite{M08}: The Hamiltonian (\ref{NHWeylHam}) is a quasi-Hermitian Hamiltonian that satisfies the relation $H\eta=\eta H^{+}$, where $\eta$ is some positive definite operator called \emph{metric operator}. The condition of the metric operator of being positive definite, allows us, among other things, to define a non-unitary similarity  transformation $S$ ($\eta=S^{+}S$) that maps the NH Hamiltonian $H$ in Eq.(\ref{NHWeylHam}) onto a \emph{hermitian} Hamiltonian $\hat{H}$ (for further details we refer to the Appendix (\ref{App:equilibrium})). Then, we can find the conserved currents in the auxiliary hermitian Hamiltonian and use the mapping $S$ on these conserved current to find the corresponding conserved quantities in the non-hermitian model.

\Al{In certain systems, where the similarity matrix $S$ cannot be constructed explicitly, it is still possible to identify certain conserved quantities. These quantities are associated with the operators that are symmetries of the system. Namely, given an operator $\mathcal{O}$, we can construct another operator $\mathcal{O}\pr=\eta\mathcal{O}$, whose time evolution is described, according to Eqs.(\ref{Heis1},\ref{TevolNH}), as follows:}
\beq
\frac{d\mathcal{O\pr}(t)}{d t}=i \eta e^{iH t}[H,\mathcal{O}]e^{-iH t}.\label{Heis2}
\eeq
\Al{We see that this new operator $\mathcal{O\pr}$ now possesses a conventional time evolution of the quantum mechanics in the Heisenberg picture. Also, if the original operator $\mathcal{O}$ is a symmetry of the problem (that is $[H,\mathcal{O}]=0$), then the new operator $\mathcal{O\pr}=\eta\mathcal{O}$ defines a conserved quantity as well. This discussion allows us to motivate the use of a bi-orthogonal formulation in our paper. It is clear, indeed, that the expression in Eq.(\ref{Heis2}), constructed with the help of the operator $\mathcal{O\pr}=\eta\mathcal{O}$, follows the standard time evolution in terms of the conjugate wavefunctions $(\bra{\psi},\ket{\psi})$.}

\Al{Alternatively, instead of using the modified operator $\mathcal{O\pr}$, we could had perfectly maintained the operator $\mathcal{O}$ and defined a modified conjugate wavefunction $\bra{\psi}\eta$. The pair $(\bra{\psi}\eta,\ket{\psi})$ is called a bi-orthogonal set. We will make use of this formulation in the next sections.}

\Al{Let us now apply the mentioned results to the model defined in Eq.(\ref{NHWeylHam}). The corresponding procedure, developed in Ref.~\cite{AMS17}, utilizes the metric operator $\eta=\mathbf{1}+\frac{m_5}{m}\gamma_5$. It turns our that the Hermitian model associated to the non-Hermitian Hamiltonian $H$ corresponds to a massive Dirac spinor $\chi$ with mass $M=\sqrt{m^2-m^2_5}$. We construct the $U(1)$ conserved current $\hat{j}^{\mu}=\overline{\chi}\gamma^{\mu}\chi$ associated with the spinor $\chi$. After using the inverted mapping $S$, we get the corresponding current in the non-Hermitian system in terms of the field $\psi^{+}$:}
\beq
j^{\mu}=\psi^{+}\gamma^{0}(\mathbf{1}+\frac{m_5}{m}\gamma_5)\gamma^{\mu}\psi=\psi^{+}\gamma^{0}\eta\gamma^{\mu}\psi=\bar{\psi}\eta\gamma^{\mu}\psi.\label{conservedcurrent}
\eeq 
\Al{Since $\partial_{\mu}\hat{j}^{\mu}=0$, can trivially show that the current $j^{\mu}$ corresponds to a conserved quantity $\partial_{\mu}j^{\mu}=0$. We thus see that the current $j^{\mu}=\eta J^{\mu}$ is a conserved current with $J^{\mu}=\gamma^\mu$ being a symmetry 
of the Hamiltonian in Eq.(\ref{NHWeylHam}).}

As it will discussed in the next section, the most important consequence of having the conserved current $j^{\mu}$ is that  we can define a chemical potential $\mu$ associated to $j^0=\eta$.

We immediately see that the current is made of a piece proportional to the identity, as it corresponds to an abelian current in the Hermitian case, together with a chiral current weighted by $m_5/m$ that implies a chiral imbalance. We will show in the next Section and in the Appendix (\ref{App:equilibrium}) that a system defined by the Hamiltonian (\ref{NHWeylHam}) that exchange particles in a manner defined by this precise chemical potential $\mu$ defines is in a trully equilibrium thermal state with non vanishing CME and CSE.

\section{Computation of CSE and CME with biorthogonal quantum mechanics} 

Here we will tackle the problem using the biorthogonal quantum mechanics formalism\cite{Br14}. Within this formalism, we distinguish between the eigenstates of $H$: $H\psi_s=\varepsilon^s_k\psi_s$, their complex conjugates: $\psi^{+}_sH^+=\psi^{+}_s\varepsilon^s_k$, the bi-orthogonal states $\phi_s$: $H^+\phi_s=\varepsilon^s_k\phi_s$, and their complex conjugates, $\phi^{+}_sH=\varepsilon^s_k\phi^+_s$. The point is that, because $H$ is not Hermitian, $\psi_s\neq \phi_s$, and $\psi^+_s\neq\phi^+_s$. Also, for the same lack of Hermititivity, the states are not orthogonal $\braket{\psi^+_s|\psi_{s\pr}}\neq \delta_{ss\pr}$, where $\braket{\cdot|\cdot}$ is the standard scalar product in the corresponding Hilbert space. However, the state sets $\psi_s$ and $\phi_s$ form a bi-orthogonal basis:

\beq
\braket{\psi^+_s|\phi_{s\pr}}=\braket{\phi^+_s|\psi_{s\pr}}\propto\delta_{ss\pr}.
\eeq
For the model (\ref{NHWeylHam}) we can define a metric operator $\eta$, that not only fulfills the quasi-Hermiticity condition, $\eta H=H^{+}\eta$ but it is positive definite. The existence of such operator simplifies the construction of the bi-orthogonal basis sets, since these two bases are related to each other through $\eta$:
\beq
\phi_s=\frac{1}{\braket{\psi^+_s|\eta\psi_s}}\eta\psi_s.
\eeq
With this particular normalization, we have $\braket{\psi^+_s|\phi_{s\pr}}=\braket{\phi^+_s|\psi_{s\pr}}=\delta_{ss\pr}$. For the Hamiltonian at hands (\ref{NHWeylHam}), such metric operator $\eta$ is $\eta=\mathbf{1}+\frac{m_5}{m}\gamma_5$\cite{AMS17}. The existence of a metric operator allowing us to define a well-defined inner product in the corresponding Hilbert space, defines an unitary time evolution of the states, as long as the spectrum is real, so a consistent description of quantum mechanics is allowed for the system, although being non-Hermitian. \Al{Also, it is now easy to see that any time operator will evolve using the conventional Heisenberg picture within the bi-orthogonal formalism.}

Another relevant consequence of the existence of the metric operator is that $\eta$ is a conserved quantity, since, as we mentioned, the matrix $\eta$ fulfills the pseudo-Hermiticity condition \Al{(remember Eq.(\ref{TevolNH}))}. Although $\eta$ does not commute with the non-Hermitian Hamiltonian $H$\cite{SBM18}, it allows for a construction of an unitary evolution. \Al{The existence of a conserved quantity makes it possible to define a Lagrange multiplier $\mu$ associated to the operator $\eta$. Since $\eta$ is a conserved quantity, that Lagrange multiplier $\mu$ plays the role of the chemical potential. Consequently, we can define a new Hamiltonian}
\beq
\mathcal{H}=H-\mu\eta,\label{HwithB}
\eeq
\Al{as it is done in the standard Hermitian statistical mechanics. Of course, due to the non-Hermitian nature of the problem, the conserved quantity does not need to commute with $H$. Instead, to be conserved, the corresponding operator should satisfies the aforementioned pseudo-Hermiticity condition. 
}

\Al{However, the existence a common basis between $H$ and any operator $\mathcal O$ is possible if and only if the operator $\mathcal O$ commutes with the Hamiltonian $H$, irrespective of the Hermiticity of $H$. This fact means that, we will not be able to find a common basis for $\eta$ and $H$ in terms of the eigenstates of the number operator, as it happens in conventional Hermitian Quantum Mechanics. This problem may be circumvented by building the bi-orthogonal basis which is, in turn, is constructed by diagonalizing the new Hamiltonian~\eq{HwithB} $\mathcal{H}$ instead of the original Hamiltonian $H$. }

\Al{
In order to compute the non-Hermitian version of the Chiral Magnetic Effect, we consider the model Eq.(\ref{HwithB}) in a classical background of an external constant magnetic field $\bm{B}$ that points along the third dimension. As it is a trivial exercise to obtain the Landau levels for this model, we are not presenting the details of the derivation. However, we are willing to highlight two properties of these Landau levels.}

\Al{
First, the algebraic wavefunction structure in the non-Hermitian system does not differ from the Hermitian case: The system is translationally invariant along the direction of the magnetic field $\bm{B}$, so that the momentum along this direction is a conserved integral of motion. Thus, we may use a standard Fourier transformation of the wavefunction along this direction. The dispersion relation of $\mathcal{H}$ for a non-zero chemical potential is presented in Fig.~\ref{fig:bands}(a).}

\Al{
Second, the wavefunctions are highly degenerate as in the Hermitian case, with the same Landau degeneracy. The current along the magnetic field becomes diagonal. Therefore, following a standard procedure used in the Hermitian case, we can integrate over the transverse spatial directions when computing quantum averages. This approach greatly simplifies the calculations and allows us to consider the problem as quasi-one dimensional.}


Although from the perspective of constructing a thermal equilibrium ensemble, the lack of Hermiticity in the system might suppose a problem\cite{RB15}, here we will argue that this is not actually the case in our model system under general grounds,  since the requirement the one-particle correlation function built from the bi-orthogonal basis satisfies the \Al{Kubo-Martin-Schwinger (KMS) periodicity condition: $\braket{\Psi^+(0)\Psi(\tau\pr)}=-\braket{\Psi^{+}(\beta)\Psi(\tau\pr)}$ \cite{K57,MS59,HHW67}. It is known in the context of quantum statistical mechanics that states satisfying the KMS boundary condition extremize the Von Neumann entropy $S=-Tr[\rho \log \rho]$, being $\rho$ the density matrix operator , so they describe equilibrium states\cite{HHW67}.} The point is to notice that, for non-zero $\mu$, the time evolution of any field operator is done through the exponential of $\mathcal{H}$ (and not of $H$) so then the one-particle correlation function will satisfy the KMS boundary condition and we will be able to built an equilibrium ensemble\cite{FW71}. \Al{In the Appendix (\ref{App:equilibrium}) we provide an explicit proof that the correlation functions of the non-Hermitian system considered in the present work can be mapped to the correlation functions of an equilibrium thermal state, thus satisfying the KMS condition.} Also, this fact has been pointed out in the existing literature of non-Hermitian systems\cite{J07,BPP20}.

\Al{In our particular case, using the effective one-dimensional model, we will focus only in the lowest Landau level (LLL) after integrating over the perpendicular coordinates and explicitly writing the Landau level denegeracy $\rho=2\pi eB_3$} the equilibrium thermal average of any observable $\mathcal{O}$ will be
\beq
\braket{\mathcal{O}}=\frac{e^2 B_3}{4\pi^2}\sum_{\omega_n}\int^{\infty}_{-\infty} d k_3 Tr[\mathcal{O}G_0(i\omega_n,k_3)],
\eeq
where $G_0(i\omega_n,k_3)$ is the single-particle propagator in imaginary time:
\beq
G_0(i\omega_n,k_3)=\sum_{s,n}\frac{\ket{\psi_s}\bra{\phi_s}}{i\omega_n-\varepsilon^{s}_{\mu}(k_3)},\label{Gfunction}
\eeq
and $(\psi_s,\phi_s)$ are the bi-orthogonal sets of single-particle eigenstates of the model (\ref{HwithB}) in presence of an external magnetic field $\bm{B}=B_3\hat{\bm{z}}$: $\mathcal{H}\psi_s=\varepsilon^{s}_{\mu}(k_3)\psi_s$, $\mathcal{H}^+\phi_s=\varepsilon^{s}_{\mu}(k_3)\phi_s$. The generic label $s$ comprises band labelling, the spin index $\tau$, and the \Al{Landau level index $N$}.

\begin{figure}
\begin{center}
\includegraphics[scale=0.23]{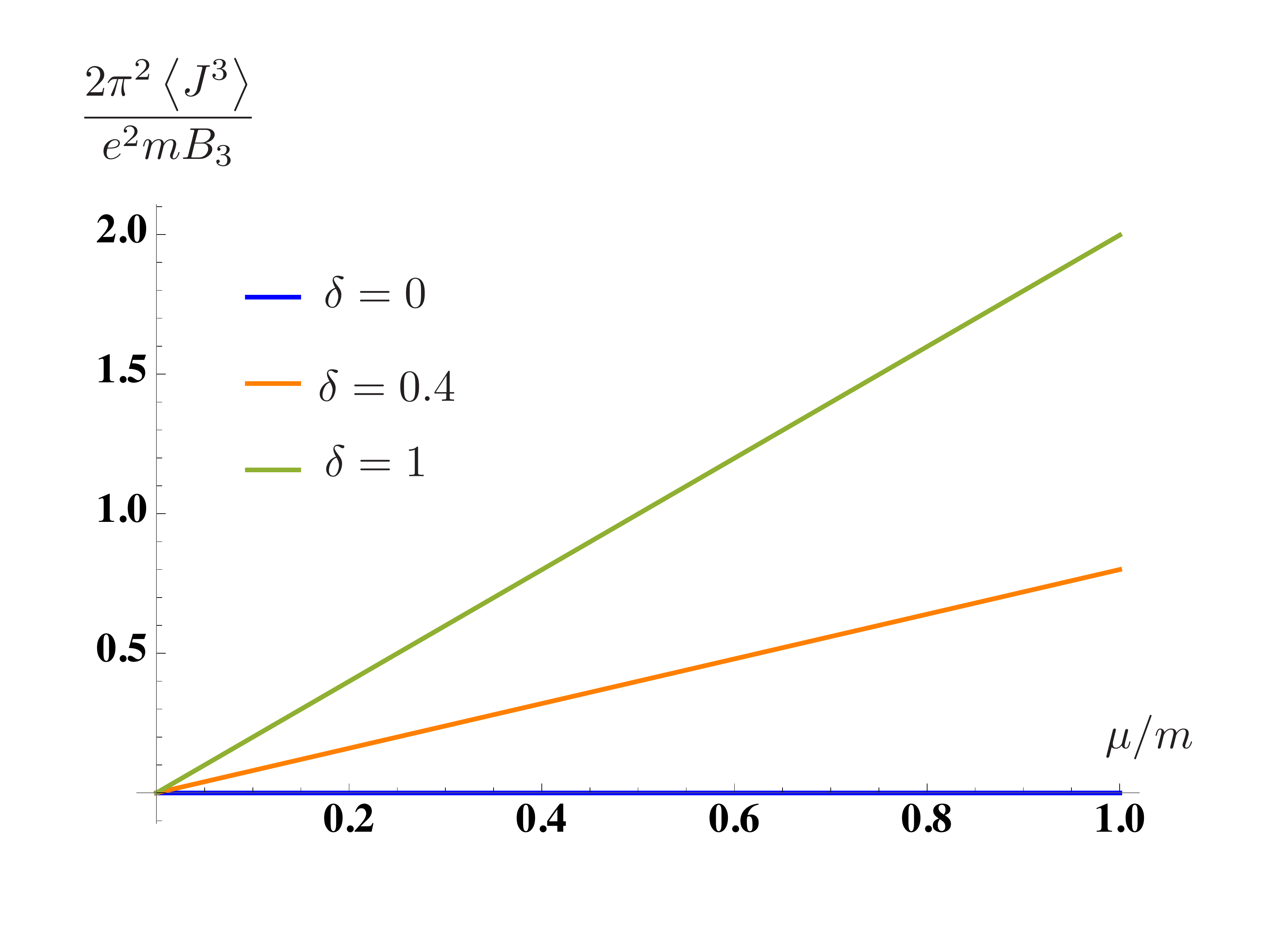}
\end{center}
\caption{(color online) CME as a function of $\mu/m$ for three values of $\delta=m_5/m$. The vanishing CME for the Hermitian case, $\delta=0$ is recovered.}
\label{fig:CMEdelta}
\end{figure}

For the operators defined as $\mathcal{O}=\frac{\partial \mathcal{H}}{\partial \lambda}$, we can generalize the Feynman-Hellmann theorem to the bi-orthogonal basis (See Appendix{\ref{App:HFtheorem}}), if the eigenstates are real: 
\beqa
\braket{\phi^+_s|\mathcal{O}\psi_s}=\braket{\phi^+_s|\frac{\partial \mathcal{H}}{\partial \lambda}\psi_s}=\frac{\partial \varepsilon_{s}}{\partial \lambda},
\eeqa

obtaining, after performing the Matsubara summation, 
\beq
\braket{\mathcal{O}}=\frac{e^2 B_3}{4\pi^2}\sum_{s,n}\int^{\infty}_{-\infty} d k_3\frac{\partial \varepsilon^{s}_{\mu}(k_3)}{\partial \lambda}n_F(\varepsilon^{s}_{\mu}(k_3)),\label{tevO}
\eeq

where $n_F(x)$ is the Fermi distribution function \emph{in absence of the chemical potential}. The chemical potential is part of the spectrum.

\begin{figure}
\begin{tabular}{cc}
\includegraphics[scale=0.2]{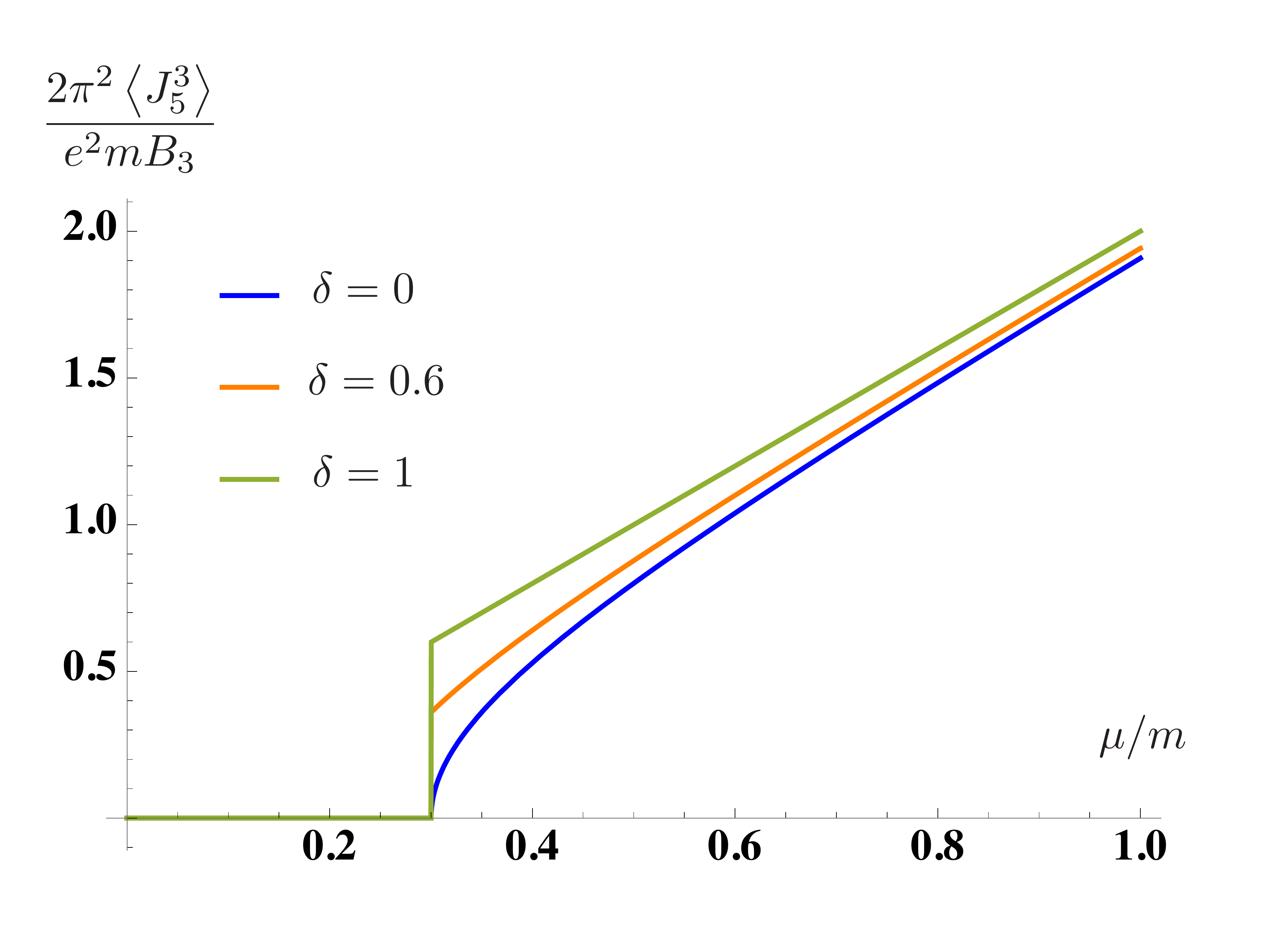} & 
\includegraphics[scale=0.2]{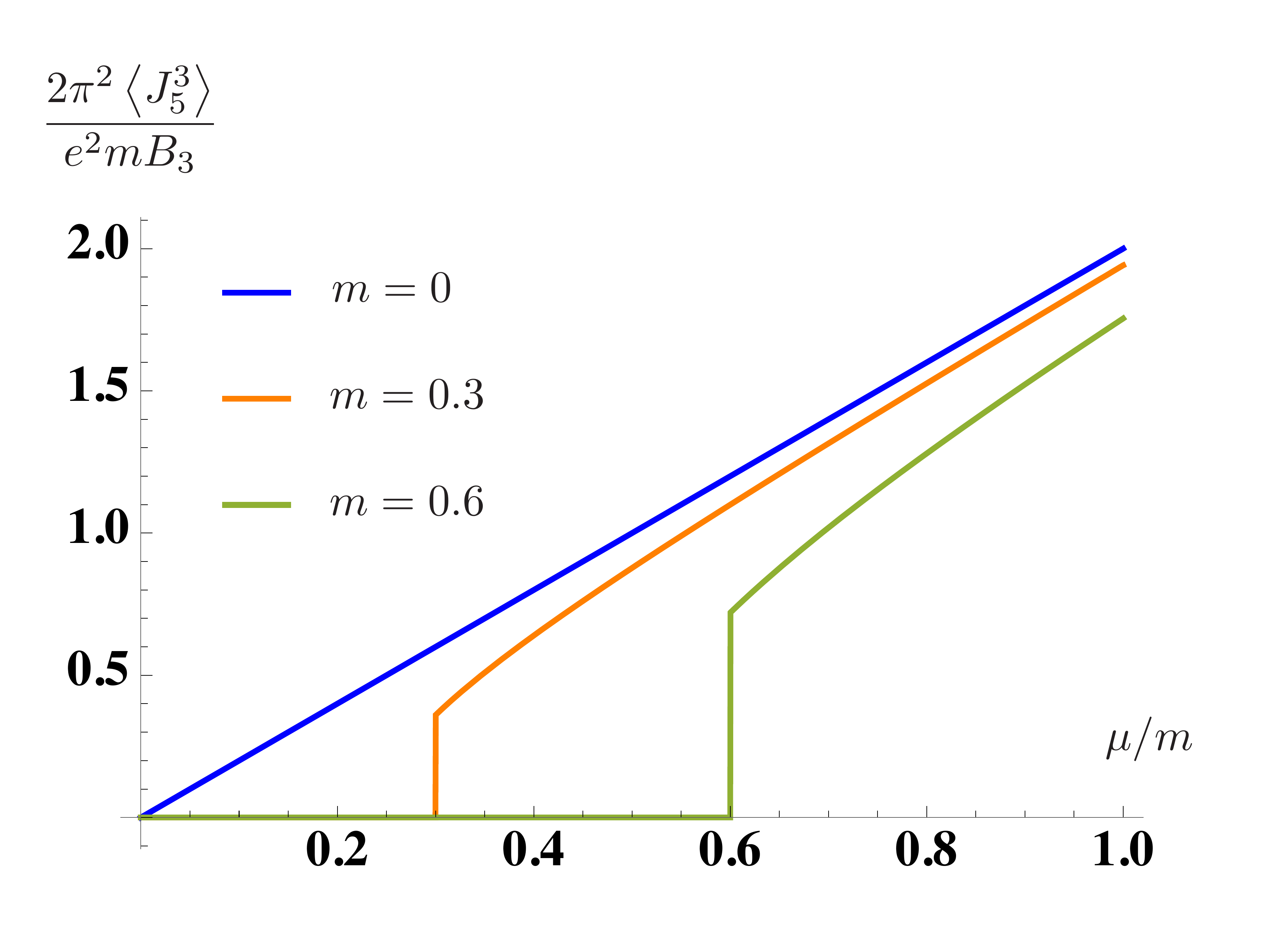} \\
(a) & (b)
\end{tabular}
\caption{(color online) (a): Regularized CSE as a function of $\mu$ for three values of $\delta=m_5/m$. We fix the mass parameter to be $m=0.3$. (b): CSE as a function of $\mu$ for three values of $m$ and fixed $\delta=0.6$.}
\label{fig:CSE}
\end{figure}

For the case of CME, $J_3=\frac{\partial \varepsilon^s_{\mu}(k_3)}{\partial k_3}$, so
\beq
\braket{J^3}=\frac{e^2B_3}{4\pi^2}\sum_{s,N}\int^{\infty}_{-\infty} d k_3\frac{\partial \varepsilon^{s,N}_{\mu}(k_3)}{\partial k_3}n_F(\varepsilon^{s,N}_{\mu}(k_3)).\label{CMEBiorthodef}
\eeq

The dispersion relation for the LLL ($N=0$) sector is (see Fig.(\ref{fig:bands}(b))):
\beq
\varepsilon^{s,0}_{\mu}(k_3)=-\mu+s\sqrt{(k_3-\delta \mu)^2+m^2(1-\delta^2)},
\eeq 
where $\delta=\frac{m_5}{m}$ and $s=\pm1$, while for $N>0$, we have
\beq
\varepsilon^{s,\tau,N}_{\mu}(k_3)=-\mu+s\sqrt{(\sqrt{k^2_3+\omega^2_c N}+\tau\delta \mu)^2+m^2(1-\delta^2)}.
\eeq

For the $N>0$ Landau levels, the spin degree of freedom $\tau=\pm1$ appears explicitly. In Fig.(\ref{fig:bands}(b)) we have plotted the Landau level spectrum for $N=0$ and $N>0$. The all important difference between the eigen-energies for $N=0$ and $N>0$ is that, while $\varepsilon^{s,\tau,n}_{\mu}(k_3)$ with $N>0$ is an even function of $k_3$ for any value of $m,\delta=m_5/m$, and $\mu$, the energy $\varepsilon^{s,0}_{\mu}(k_3)$ with $N=0$ is not. That means that, when taking the derivative with respect to $k_3$ and integrating over a symmetric interval, the $N>0$ Landau levels will not contribute to the integral in (\ref{CMEBiorthodef}), but the $N=0$ will do.

The result turns out to be (see Fig.(\ref{fig:CMEdelta})):
\beq
\braket{J^3}=\frac{e^2B_3}{2\pi^2}\frac{m_5}{m}\mu.\label{CMEBiortho}
\eeq

This is the principal result of this Letter. For non-zero values of the mass $m_5$, which is the parameter that controls the non-Hermiticity of $\mathcal{H}$, there is a \emph{non-vanishing CME in equilibrium}.

The chiral separation effect (CSE) is obtained by computing the average value of the chiral current, represented by the operator $J^{i}_5=e\alpha^i\gamma_5$.  We can follow the same route as in the case of the CME. We will add a term $b_3\alpha_3\gamma_5$ to the Hamiltonian (\ref{HwithB}) and compute the spectrum in presence of the parameter $b_3$. Then, we apply the Hellman-Feynman theorem to it, taking the derivative with respect to $b_3$ and constructing the expectation value for each Landau level. We send the parameter $b_3$ to zero after the calculation. 
It is a lengthy but straightforward calculation to check that for the $n>0$ sector, $\partial \varepsilon^{s,n}_\mu(k_3,b_3)/\partial b_3$ is an odd function of $k_3$ in the limit $b_3\to 0$ for all values of $m$, $m_5$, and $\mu$. This implies that the integral over $k_3$ is zero and they do not contribute to the CSE. In contrast,  for the $n=0$ sector, we simply have
$\partial \varepsilon^{s}_{\mu}(k_3,b_3)/\partial b_3=1$,
so
\beq
\braket{J^3_5}=\frac{e^2 B_3}{4\pi^2}\sum_{s}\int^{\infty}_{-\infty} d k_3 n_F(\varepsilon^{s,0}_{\mu}(k_3,b_3=0)).\label{CSEBiorthodef}
\eeq
We have plotted $\braket{J^3_5}$ in Figs.(\ref{fig:CSE}a,\ref{fig:CSE}b) as a function of $\delta=m_5/m$ for fixed $m$, and as function of $m$ for fixed $\delta$.
Performing the integral we finally have
\beq
\braket{J^3_5}=\frac{e^2 B_3}{2\pi^2}\left(\frac{1}{\epsilon}+\Theta[\mu-m]\sqrt{\mu^2-m^2(1-\delta^2)}\right),\label{CSEBiortho}
\eeq
with $\epsilon\ll1$. We note that there is a divergent contribution in the CSE. It is a particular feature of $(1+1)$ dimensions that there is a duality between the chiral and charge currents. The charge current operator representing the CME is the chiral density, while the chiral current operator $j^5_1$ that is relevant for the study of the CSE is the same operator as the charge density. Having the same origin as the standard charge density, we regularize it in the same way.

\section{Conclusions} 

In the present Letter we have demonstrated that CME in equilibrium is possible when non-Hermitian systems are considered. The key ingredient is to realize that the CME is zero if charge conservation is imposed in the system. However, charge conservation, associated to the $U(1)$ symmetry, is not fulfilled in non-Hermitian systems as it is done in conventional Hermitian ones. 

Another fact to pay attention is that there not an unique metric operator associated to the non-Hermitian Hamiltonian fulfilling the pseudo-Hermiticity condition. While there is no practical consequence of this regarding the construction of a bi-orthogonal basis (the average of observables do not depend on any particular choice of the metric operator), this observation is relevant as we can associate different chemical potentials to different metric operators understood as conserved quantities in the non-Hermitian sense. Interestingly, all the metric operators are related to each other by a similarity transformation\cite{M08}, so we can generalize the results obtained here to other chemical potentials by modifying the spectrum correspondingly.

Finally, to the best of our knowledge, there are not experimental realizations of fermionic non-Hermitian systems with real spectrum to test our predictions. However, there are impressive experimental advances in the area of non-Hermitian $\mathcal{PT}-$symmetric photonic systems and other condensed matter analogs\cite{ZVP14,H18}. In fact, it has been recently proposed the experimental observation of the CME employing superconducting quantum circuit technology, and synthetic magnetic fields\cite{TZL18}. We suggest the same experimental setup to test our theory, by extending the experimental setup with equal gain-loss\cite{QNO18}. 
Besides, other topological equilibrium effects similar to the CME have been proposed to occur in electromagnetism\cite{AS17,Y17,ARN17,CCL18}, being the optical helicity and the optical chirality the electromagnetic symmetries that play the role of the chiral symmetry in ultrarelativistic fermionic systems. There, the biorthogonal formalism have probed to be useful to handle the effect of dissipation and loss in electromagnetism\cite{ABN18,VEM18}. The natural question is then to see how the topologically-related responses associated to these symmetries are modified by the presence of non-Hermitian effects.

\section{Acknowledgements} 

We kindly acknowledge inspiring conversations with K. Landsteiner about non-Hermitian systems and the physics underlying the CME.
A.C. acknowledges financial support through the MINECO/AEI/FEDER, UE Grant No. FIS2015-73454-JIN. and the Comunidad de Madrid MAD2D-CM Program (S2013/MIT-3007), and the Ramon y Cajal program through the grant RYC2018-023938-I. The research of M.C. was partially supported by Grant No. 0657-2020-0015 of the Ministry of Science and Higher Education of Russia.

 
\appendix
\appendixtitles{yes}
\section{The Hellman-Feynman theorem for bi-orthogonal systems}\label{App:HFtheorem}

In this Appendix we give a proof of the extension of the Hellman-Feynman theorem to bi-orthogonal systems with real spectrum. 

As discussed in the main text, the bi-orthogonal basis is constructed with two set of states satisfying $H\ket{n}=\varepsilon_n\ket{n}$, $H^+\ket{\overline{n}}=\varepsilon_n\ket{\overline{n}}$, and $\bra{n}H^+=\bra{n}\varepsilon_n$, $\bra{\overline{n}}H=\bra{\overline{n}}\varepsilon_n$, together with the normalization condition $\braket{\overline{n}|n\pr}=\braket{n\pr|\overline{n}}=\delta_{nn\pr}$.

Let us consider a Hamiltonian $H$ depending on some parameter $\lambda$. To ease notation, we will keep the dependence with the generic parameter $\lambda$ implicit in the eigenstates and eigenvalues. We are interested in computing the averaged value 
\beq
\braket{\overline{n}|\frac{\partial H}{\partial \lambda}|n}.
\eeq

Then, we compute
\beqa
&&\frac{\partial}{\partial \lambda}\braket{\overline{n}|H|n}=\frac{\partial}{\partial \lambda}(\varepsilon_n\braket{\overline{n}|n})=\nonumber\\
&=&\braket{\frac{\partial}{\partial \lambda}\overline{n}|H|n}+\braket{\overline{n}|H|\frac{\partial}{\partial \lambda}n}+
\braket{\overline{n}|\frac{\partial H}{\partial \lambda}|n}=\nonumber\\
&=&\varepsilon_n\frac{\partial}{\partial \lambda}(\braket{\overline{n}|n})+\braket{\overline{n}|\frac{\partial H}{\partial \lambda}|n}.
\eeqa
As $\ket{s}$ and $\ket{\overline{n}}$ are eigenstates of $H$ and $H^+$ with the same eigenvalue $\varepsilon_n$.
Simplifying a little, we finally have:
\beqa
\frac{\partial \varepsilon_n}{\partial \lambda}=\braket{\overline{n}|\frac{\partial H}{\partial \lambda}|n},
\eeqa
which is the result we wanted to prove.

\section{Thermal equilibrium condition in quasi-Hermitian systems}
\label{App:equilibrium}

For Hermitian systems, the condition of thermal equilibrium can be formally stablished by showing that the Hermitian system satisfies the Kubo-Martin-Schwinger (KMS) boundary condition for the imaginary-time propagator\cite{HHW67}. For quasi-Hermitian systems, it is possible to describe equilibrium in the same way, making use of the existing non-unitary mapping between the non-Hermitian and Hermitian Hamiltonians. In what follows, we will restrict ourselves to non-Hermitian systems described by Hamiltonian operators that do not depend on time.

Let us consider two operators $A(\tau)$ and $B(\tau)$ in the Heisenberg picture (and in the imaginary time formalism) described by the Hamiltonian $\mathcal{H}$ (we consider that chemical potentials associated to symmetries of the problem are already included in $\mathcal{H}$). The KSM condition can be stated as the following identity:
\beq
Tr[e^{-\beta\mathcal{H}}A(\tau)B(\tau\pr)]=Tr[e^{-\beta\mathcal{H}}B(\tau\pr)A(\tau+\beta)].
\eeq
If $A=\psi^+$ and $B=\psi$ are field operators that anticommute, we have
\beq
Tr[e^{-\beta\mathcal{H}}\psi^+(0)\psi(\tau\pr))]=-Tr[e^{-\beta\mathcal{H}}\psi^+(\beta)\psi(\tau\pr)].\label{KSM}
\eeq
As explained in \cite{FW71}, this means that the thermal averaged propagator $\braket{T \psi^+(\tau)\psi(\tau\pr)}$ ($T$ refers to the Dyson time ordering) is an antiperiodic function of $\tau$ with period $\beta$. This allows the development of all the machinery of thermal field theory.

In order to show how this works for quasi-Hermitian systems, it is enough to show that, for a quasi-Hermitian Hamiltonian, it is possible to construct an Hermitian partner through a non-unitary mapping between them, so we map the statistical averages using the bi-orthogonal basis in the non-Hermitian case, map them to their Hermitian counterparts, establish the KSM condition in the latter, and going back to the non-Hermitian case inverting the mentioned mapping.

As demonstrated in \cite{J07}, the existence of a metric operator $\eta$ allows us to define the non-unitary mapping $S$ of some quasi-Hermitian Hamiltonian $\mathcal{H}$ to an Hermitian partner $\hat{\mathcal{H}}$, with $\hat{\mathcal{H}}=\hat{\mathcal{H}}^+$ (we will denote the Hermitian partners of operators with the hat symbol $\hat{ }$ ):
\beq
\hat{\mathcal{H}}=S\mathcal{H}S^{-1}.
\eeq
Also, we can define the Hermitian partner of any operator associated to the quasi-Hermitian system in the same way:

\beq
\hat{\mathcal{O}}=S\mathcal{O}S^{-1}.
\eeq
This includes the field operators $\Psi$ and $\Psi^+$ in the second quantization formalism. As discussed in the main text, the existence of the metric operator $\eta$ allows us to construct a well behaved scalar product in the Hilbert space and to construct bi-orthogonal basis sets, ${\ket{n}}$ and ${\ket{\overline{n}}}$. In this way, we can define the following statistical average (here we will use the suffix $bi$ to denote the statistical average with the bi-orthogonal basis):

\beqa
&&\braket{\mathcal{O}}_{bi}\equiv\sum_n\braket{\phi^+_{n}  e^{-\beta\mathcal{H}}\mathcal{O}\psi_n}=
\sum_n\frac{1}{\braket{\psi^+_n\eta\psi_n}}\braket{\psi^+_{n} \eta e^{-\beta\mathcal{H}}\mathcal{O}\psi_n}=\nonumber\\
&=&\sum_n\frac{1}{\braket{\hat{\psi}^+_n \hat{\psi}_n}}\braket{\hat{\psi}^+_n\underbrace{S^{-1}S}_{\mathbf{1}} \underbrace{S S^{-1}}_{\mathbf{1}}e^{-\beta\hat{\mathcal{H}}}\underbrace{SS^{-1}}_{\mathbf{1}}\hat{\mathcal{O}}\underbrace{S S^{-1}}_{\mathbf{1}}\hat{\psi}_n}=\nonumber\\
&=&\sum_n\braket{\hat{\psi}^+_n e^{-\beta\hat{\mathcal{H}}}\hat{\mathcal{O}}\hat{\psi}_n}=\braket{\hat{\mathcal{O}}}.\label{identity}
\eeqa
In the second line we have used $\eta=S S$ ($S^+=S$ in our particular case), and that the eigenstates of the non-Hermitian $\mathcal{H}$ are related to the eigenstates of the Hermitian partner $\hat{\mathcal{H}}$ through $\hat{\psi}_n=S\psi_n$. Also, we consider that the states $\psi_n$ of the Hermitian partner are conveniently normalized:$\braket{\hat{\psi}^+_n \hat{\psi}_n}=1$ .

To guarantee the proper normalization of  (\ref{identity}), we need to relate the partition function in the quasi-Hermitian system and its Hermitian partner. This is a particular case of the previous identity, as we can choose $\mathcal{O}=\mathbf{1}$ and obtain the equality of the corresponding partition functions:

\beqa
&&Z_{bi}=\sum_n\braket{\phi^+_n e^{-\beta\mathcal{H}}\psi_n}=
\sum_n\frac{1}{\braket{\psi^+_n\eta\psi_n}}\braket{\psi^+_n \eta e^{-\beta\mathcal{H}}\psi_n}=\nonumber\\
&=&\sum_n\frac{1}{\braket{\hat{\psi}^+_n \hat{\psi}_n}}\braket{\hat{\psi}^+_n \underbrace{S^{-1}S}_{\mathbf{1}}\underbrace{S S^{-1}}_{\mathbf{1}}e^{-\beta\hat{\mathcal{H}}}\underbrace{S S^{-1}}_{\mathbf{1}}\hat{\psi}_n}=\nonumber\\
&=&\sum_n\braket{\hat{\psi}^+_n e^{-\beta\hat{\mathcal{H}}}\hat{\psi}_n}=
\hat{Z},
\eeqa
where we have denoted the partition function of the Hermitian partner by $\hat{Z}$.

We can generalize (\ref{identity}) to any product of field operators. Then we obtain that

\beqa
&&\braket{\Psi^+(0)\Psi(\tau\pr)}_{bi}=\braket{\hat{\Psi}^+(0)\hat{\Psi}(\tau\pr)}=\nonumber\\
&=&-\braket{\hat{\Psi}^+(\beta)\hat{\Psi}(\tau\pr)}=-\braket{\Psi(\beta)^{+}\Psi(\tau\pr)}_{bi},
\eeqa
so we conclude that the averages performed with the bi-orthogonal basis and with the density matrix $\rho=e^{-\beta\mathcal{H}}$ satisfy a KMS boundary condition and thus this state defines a thermal state in equilibrium, since it is trivial to modify the previous reasoning by including the Dyson time ordering operation. Also, this reasoning justifies the definition of the discrete-frequency Green function $G_{0}(i\omega_n)$ in Eq.(\ref{Gfunction}) of the main text.



\externalbibliography{yes}
\bibliography{NHCME_Symmetry_v2}

\begin{thebibliography}{-------}
\providecommand{\natexlab}[1]{#1}

\bibitem[Fukushima \em{et~al.}(2008)Fukushima, Kharzeev, and Warringa]{FKW08}
Fukushima, K.; Kharzeev, D.E.; Warringa, H.J.
\newblock Chiral magnetic effect.
\newblock {\em Phys. Rev. D} {\bf 2008}, {\em 78},~074033.
\newblock
  doi:{\changeurlcolor{black}\href{https://doi.org/10.1103/PhysRevD.78.074033}{\detokenize{10.1103/PhysRevD.78.074033}}}.

\bibitem[Adler(1969)]{A69}
Adler, S.L.
\newblock Axial-Vector Vertex in Spinor Electrodynamics.
\newblock {\em Phys. Rev.} {\bf 1969}, {\em 177},~2426--2438.
\newblock
  doi:{\changeurlcolor{black}\href{https://doi.org/10.1103/PhysRev.177.2426}{\detokenize{10.1103/PhysRev.177.2426}}}.

\bibitem[Bell and Jackiw(1969)]{BJ69}
Bell, J.S.; Jackiw, R.
\newblock A PCAC puzzle: ?0???in the ?-model.
\newblock {\em Il Nuovo Cimento A (1965-1970)} {\bf 1969}, {\em 60},~47--61.
\newblock
  doi:{\changeurlcolor{black}\href{https://doi.org/10.1007/BF02823296}{\detokenize{10.1007/BF02823296}}}.

\bibitem[Kharzeev(2014)]{Kharzeev:2013ffa}
Kharzeev, D.E.
\newblock {The Chiral Magnetic Effect and Anomaly-Induced Transport}.
\newblock {\em Prog. Part. Nucl. Phys.} {\bf 2014}, {\em 75},~133--151,
  \href{http://xxx.lanl.gov/abs/1312.3348}{{\normalfont
  [arXiv:hep-ph/1312.3348]}}.
\newblock
  doi:{\changeurlcolor{black}\href{https://doi.org/10.1016/j.ppnp.2014.01.002}{\detokenize{10.1016/j.ppnp.2014.01.002}}}.

\bibitem[Landsteiner(2016)]{L16}
Landsteiner, K.
\newblock Notes on anomaly induced transport.
\newblock {\em Acta Phys. Pol B} {\bf 2016}, {\em 47},~2617.

\bibitem[Ma and Pesin(2015)]{MP15}
Ma, J.; Pesin, D.A.
\newblock Chiral magnetic effect and natural optical activity in metals with or
  without Weyl points.
\newblock {\em Phys. Rev. B} {\bf 2015}, {\em 92},~235205.
\newblock
  doi:{\changeurlcolor{black}\href{https://doi.org/10.1103/PhysRevB.92.235205}{\detokenize{10.1103/PhysRevB.92.235205}}}.

\bibitem[Yamamoto(2015)]{Y15}
Yamamoto, N.
\newblock Generalized Bloch theorem and chiral transport phenomena.
\newblock {\em Phys. Rev. D} {\bf 2015}, {\em 92},~085011.
\newblock
  doi:{\changeurlcolor{black}\href{https://doi.org/10.1103/PhysRevD.92.085011}{\detokenize{10.1103/PhysRevD.92.085011}}}.

\bibitem[Zubkov(2016)]{Z16}
Zubkov, M.A.
\newblock Absence of equilibrium chiral magnetic effect.
\newblock {\em Phys. Rev. D} {\bf 2016}, {\em 93},~105036.
\newblock
  doi:{\changeurlcolor{black}\href{https://doi.org/10.1103/PhysRevD.93.105036}{\detokenize{10.1103/PhysRevD.93.105036}}}.

\bibitem[Bohm(1949)]{B49}
Bohm, D.
\newblock Note on a Theorem of Bloch Concerning Possible Causes of
  Superconductivity.
\newblock {\em Phys. Rev.} {\bf 1949}, {\em 75},~502--504.
\newblock
  doi:{\changeurlcolor{black}\href{https://doi.org/10.1103/PhysRev.75.502}{\detokenize{10.1103/PhysRev.75.502}}}.

\bibitem[Nielsen and Ninomiya(1981{\natexlab{a}})]{NN181}
Nielsen, H.; Ninomiya, M.
\newblock Absence of neutrinos on a lattice: (I). Proof by homotopy theory.
\newblock {\em Nuclear Physics B} {\bf 1981}, {\em 185},~20 -- 40.
\newblock
  doi:{\changeurlcolor{black}\href{https://doi.org/https://doi.org/10.1016/0550-3213(81)90361-8}{\detokenize{https://doi.org/10.1016/0550-3213(81)90361-8}}}.

\bibitem[Nielsen and Ninomiya(1981{\natexlab{b}})]{NN281}
Nielsen, H.; Ninomiya, M.
\newblock Absence of neutrinos on a lattice: (II). Intuitive topological proof.
\newblock {\em Nuclear Physics B} {\bf 1981}, {\em 193},~173 -- 194.
\newblock
  doi:{\changeurlcolor{black}\href{https://doi.org/https://doi.org/10.1016/0550-3213(81)90524-1}{\detokenize{https://doi.org/10.1016/0550-3213(81)90524-1}}}.

\bibitem[O'Brien \em{et~al.}(2017)O'Brien, Beenakker, and Adagideli]{OBA17}
O'Brien, T.E.; Beenakker, C.W.J.; Adagideli, I.
\newblock Superconductivity Provides Access to the Chiral Magnetic Effect of an
  Unpaired Weyl Cone.
\newblock {\em Phys. Rev. Lett.} {\bf 2017}, {\em 118},~207701.
\newblock
  doi:{\changeurlcolor{black}\href{https://doi.org/10.1103/PhysRevLett.118.207701}{\detokenize{10.1103/PhysRevLett.118.207701}}}.

\bibitem[Bender(2005)]{B05}
Bender, C.M.
\newblock NON-HERMITIAN QUANTUM FIELD THEORY.
\newblock {\em International Journal of Modern Physics A} {\bf 2005}, {\em
  20},~4646--4652.
\newblock
  doi:{\changeurlcolor{black}\href{https://doi.org/10.1142/S0217751X05028326}{\detokenize{10.1142/S0217751X05028326}}}.

\bibitem[Chernodub(2017)]{C17}
Chernodub, M.N.
\newblock The Nielsen-Ninomiya theorem, $PT$-invariant non-Hermiticity and
  single 8-shaped Dirac cone.
\newblock {\em Journal of Physics A: Mathematical and Theoretical} {\bf 2017},
  {\em 50},~385001.

\bibitem[Gong \em{et~al.}(2018)Gong, Ashida, Kawabata, Takasan, Higashikawa,
  and Ueda]{GAK18}
Gong, Z.; Ashida, Y.; Kawabata, K.; Takasan, K.; Higashikawa, S.; Ueda, M.
\newblock Topological Phases of Non-Hermitian Systems.
\newblock {\em Phys. Rev. X} {\bf 2018}, {\em 8},~031079.
\newblock
  doi:{\changeurlcolor{black}\href{https://doi.org/10.1103/PhysRevX.8.031079}{\detokenize{10.1103/PhysRevX.8.031079}}}.

\bibitem[Gardas \em{et~al.}(2016)Gardas, Deffner, and Saxena]{GDS16}
Gardas, B.; Deffner, S.; Saxena, A.
\newblock Non-hermitian quantum thermodynamics.
\newblock {\em Scientific Reports} {\bf 2016}, {\em 6},~23408 EP --.

\bibitem[Alexandre and Bender(2015)]{AB15}
Alexandre, J.; Bender, C.M.
\newblock Foldy-Wouthuysen transformation for non-Hermitian Hamiltonians.
\newblock {\em Journal of Physics A: Mathematical and Theoretical} {\bf 2015},
  {\em 48},~185403.

\bibitem[Alexandre \em{et~al.}(2015)Alexandre, Bender, and Millington]{ABM15}
Alexandre, J.; Bender, C.M.; Millington, P.
\newblock Non-Hermitian extension of gauge theories and implications for
  neutrino physics.
\newblock {\em Journal of High Energy Physics} {\bf 2015}, {\em 2015},~111.
\newblock
  doi:{\changeurlcolor{black}\href{https://doi.org/10.1007/JHEP11(2015)111}{\detokenize{10.1007/JHEP11(2015)111}}}.

\bibitem[Alexandre \em{et~al.}(2017)Alexandre, Millington, and Seynaeve]{AMS17}
Alexandre, J.; Millington, P.; Seynaeve, D.
\newblock Symmetries and conservation laws in non-Hermitian field theories.
\newblock {\em Phys. Rev. D} {\bf 2017}, {\em 96},~065027.
\newblock
  doi:{\changeurlcolor{black}\href{https://doi.org/10.1103/PhysRevD.96.065027}{\detokenize{10.1103/PhysRevD.96.065027}}}.

\bibitem[Armitage \em{et~al.}(2018)Armitage, Mele, and Vishwanath]{AMV18}
Armitage, N.P.; Mele, E.J.; Vishwanath, A.
\newblock Weyl and Dirac semimetals in three-dimensional solids.
\newblock {\em Rev. Mod. Phys.} {\bf 2018}, {\em 90},~015001.
\newblock
  doi:{\changeurlcolor{black}\href{https://doi.org/10.1103/RevModPhys.90.015001}{\detokenize{10.1103/RevModPhys.90.015001}}}.

\bibitem[{Mannheim}(2018)]{M18}
{Mannheim}, P.D.
\newblock {Goldstone bosons and the Higgs mechanism in non-Hermitian theories}.
\newblock {\em arXiv e-prints} {\bf 2018}, p. arXiv:1808.00437,
  \href{http://xxx.lanl.gov/abs/1808.00437}{{\normalfont
  [arXiv:hep-th/1808.00437]}}.

\bibitem[Graefe \em{et~al.}(2010)Graefe, H\:{o}ning, and Korsch]{GHK10}
Graefe, E.M.; H\:{o}ning, M.; Korsch, H.J.
\newblock Classical limit of non-Hermitian quantum dynamics{\textemdash}a
  generalized canonical structure.
\newblock {\em Journal of Physics A: Mathematical and Theoretical} {\bf 2010},
  {\em 43},~075306.
\newblock
  doi:{\changeurlcolor{black}\href{https://doi.org/10.1088/1751-8113/43/7/075306}{\detokenize{10.1088/1751-8113/43/7/075306}}}.

\bibitem[Sergi and Zloshchastiev(2013)]{SZ13}
Sergi, A.; Zloshchastiev, K.G.
\newblock Non-Hermitian Quantum Dynamics of a Two-level System and Models of
  Dissipative Environments.
\newblock {\em International Journal of Modern Physics B} {\bf 2013}, {\em
  27},~1350163,
  \href{http://xxx.lanl.gov/abs/https://doi.org/10.1142/S0217979213501634}{{\normalfont
  [https://doi.org/10.1142/S0217979213501634]}}.
\newblock
  doi:{\changeurlcolor{black}\href{https://doi.org/10.1142/S0217979213501634}{\detokenize{10.1142/S0217979213501634}}}.

\bibitem[Mostafazadeh(2008)]{M08}
Mostafazadeh, A.
\newblock Metric operators for quasi-Hermitian Hamiltonians and symmetries of
  equivalent Hermitian Hamiltonians.
\newblock {\em Journal of Physics A: Mathematical and Theoretical} {\bf 2008},
  {\em 41},~244017.

\bibitem[Brody(2014)]{Br14}
Brody, D.C.
\newblock Biorthogonal quantum mechanics.
\newblock {\em Journal of Physics A: Mathematical and Theoretical} {\bf 2014},
  {\em 47},~035305.

\bibitem[Simon \em{et~al.}(2018)Simon, Buendia, and Muga]{SBM18}
Simon, M.A.; Buendia, A.; Muga, J.G.
\newblock Symmetries and Invariants for Non-Hermitian Hamiltonians.
\newblock {\em Mathematics} {\bf 2018}, {\em 6}.
\newblock
  doi:{\changeurlcolor{black}\href{https://doi.org/10.3390/math6070111}{\detokenize{10.3390/math6070111}}}.

\bibitem[Rotter and Bird(2015)]{RB15}
Rotter, I.; Bird, J.P.
\newblock A review of progress in the physics of open quantum systems: theory
  and experiment.
\newblock {\em Reports on Progress in Physics} {\bf 2015}, {\em 78},~114001.
\newblock
  doi:{\changeurlcolor{black}\href{https://doi.org/10.1088/0034-4885/78/11/114001}{\detokenize{10.1088/0034-4885/78/11/114001}}}.

\bibitem[Kubo(1957)]{K57}
Kubo, R.
\newblock Statistical-Mechanical Theory of Irreversible Processes. I. General
  Theory and Simple Applications to Magnetic and Conduction Problems.
\newblock {\em Journal of the Physical Society of Japan} {\bf 1957}, {\em
  12},~570--586.
\newblock
  doi:{\changeurlcolor{black}\href{https://doi.org/10.1143/JPSJ.12.570}{\detokenize{10.1143/JPSJ.12.570}}}.

\bibitem[Martin and Schwinger(1959)]{MS59}
Martin, P.C.; Schwinger, J.
\newblock Theory of Many-Particle Systems. I.
\newblock {\em Phys. Rev.} {\bf 1959}, {\em 115},~1342--1373.
\newblock
  doi:{\changeurlcolor{black}\href{https://doi.org/10.1103/PhysRev.115.1342}{\detokenize{10.1103/PhysRev.115.1342}}}.

\bibitem[Haag \em{et~al.}(1967)Haag, Hugenholtz, and Winnink]{HHW67}
Haag, R.; Hugenholtz, N.M.; Winnink, M.
\newblock On the equilibrium states in quantum statistical mechanics.
\newblock {\em Communications in Mathematical Physics} {\bf 1967}, {\em
  5},~215--236.

\bibitem[Fetter and Walecka(1971)]{FW71}
Fetter, A.L.; Walecka, J.D.
\newblock {\em Quantum Theory of Many-Particle Systems}; McGraw-Hill,  1971.

\bibitem[Jakubsky(2007)]{J07}
Jakubsky, V.
\newblock THERMODYNAMICS OF PSEUDO-HERMITIAN SYSTEMS IN EQUILIBRIUM.
\newblock {\em Mod. Phys. Lett. A} {\bf 2007}, {\em 22},~1075--1084.
\newblock
  doi:{\changeurlcolor{black}\href{https://doi.org/10.1142/S0217732307023419}{\detokenize{10.1142/S0217732307023419}}}.

\bibitem[Bebiano \em{et~al.}(2020)Bebiano, da~Providencia, and
  da~Providencia]{BPP20}
Bebiano, N.; da~Providencia, J.; da~Providencia, J.P.
\newblock Toward non-Hermitian quantum statistical thermodynamics.
\newblock {\em Journal of Mathematical Physics} {\bf 2020}, {\em 61},~022102.
\newblock
  doi:{\changeurlcolor{black}\href{https://doi.org/10.1063/1.5122182}{\detokenize{10.1063/1.5122182}}}.

\bibitem[Zyablovsky \em{et~al.}(2014)Zyablovsky, Vinogradov, Pukhov,
  Dorofeenko, and Lisyansky]{ZVP14}
Zyablovsky, A.A.; Vinogradov, A.P.; Pukhov, A.A.; Dorofeenko, A.V.; Lisyansky,
  A.A.
\newblock PT-symmetry in optics.
\newblock {\em Physics-Uspekhi} {\bf 2014}, {\em 57},~1063.

\bibitem[Hayata(2018)]{H18}
Hayata, T.
\newblock Chiral magnetic effect of light.
\newblock {\em Phys. Rev. B} {\bf 2018}, {\em 97},~205102.
\newblock
  doi:{\changeurlcolor{black}\href{https://doi.org/10.1103/PhysRevB.97.205102}{\detokenize{10.1103/PhysRevB.97.205102}}}.

\bibitem[{Tan} \em{et~al.}(2018){Tan}, {Zhao}, {Liu}, {Xue}, {Yu}, {Wang}, and
  {Yu}]{TZL18}
{Tan}, X.; {Zhao}, Y.; {Liu}, Q.; {Xue}, G.; {Yu}, H.; {Wang}, Z.; {Yu}, Y.
\newblock {Emulating topological chiral magnetic effects in artificial Weyl
  semimetals}.
\newblock {\em arXiv e-prints} {\bf 2018}, p. arXiv:1802.08371,
  \href{http://xxx.lanl.gov/abs/1802.08371}{{\normalfont
  [arXiv:cond-mat.mes-hall/1802.08371]}}.

\bibitem[Quijandr\'{\i}a \em{et~al.}(2018)Quijandr\'{\i}a, Naether, \"Ozdemir,
  Nori, and Zueco]{QNO18}
Quijandr\'{\i}a, F.; Naether, U.; \"Ozdemir, S.K.; Nori, F.; Zueco, D.
\newblock $\mathcal{PT}$-symmetric circuit QED.
\newblock {\em Phys. Rev. A} {\bf 2018}, {\em 97},~053846.
\newblock
  doi:{\changeurlcolor{black}\href{https://doi.org/10.1103/PhysRevA.97.053846}{\detokenize{10.1103/PhysRevA.97.053846}}}.

\bibitem[Avkhadiev and Sadofyev(2017)]{AS17}
Avkhadiev, A.; Sadofyev, A.V.
\newblock Chiral vortical effect for bosons.
\newblock {\em Phys. Rev. D} {\bf 2017}, {\em 96},~045015.
\newblock
  doi:{\changeurlcolor{black}\href{https://doi.org/10.1103/PhysRevD.96.045015}{\detokenize{10.1103/PhysRevD.96.045015}}}.

\bibitem[Yamamoto(2017)]{Y17}
Yamamoto, N.
\newblock Photonic chiral vortical effect.
\newblock {\em Phys. Rev. D} {\bf 2017}, {\em 96},~051902.
\newblock
  doi:{\changeurlcolor{black}\href{https://doi.org/10.1103/PhysRevD.96.051902}{\detokenize{10.1103/PhysRevD.96.051902}}}.

\bibitem[Agullo \em{et~al.}(2017)Agullo, del Rio, and Navarro-Salas]{ARN17}
Agullo, I.; del Rio, A.; Navarro-Salas, J.
\newblock Electromagnetic Duality Anomaly in Curved Spacetimes.
\newblock {\em Phys. Rev. Lett.} {\bf 2017}, {\em 118},~111301.
\newblock
  doi:{\changeurlcolor{black}\href{https://doi.org/10.1103/PhysRevLett.118.111301}{\detokenize{10.1103/PhysRevLett.118.111301}}}.

\bibitem[Chernodub \em{et~al.}(2018)Chernodub, Cortijo, and Landsteiner]{CCL18}
Chernodub, M.N.; Cortijo, A.; Landsteiner, K.
\newblock Zilch vortical effect.
\newblock {\em Phys. Rev. D} {\bf 2018}, {\em 98},~065016.
\newblock
  doi:{\changeurlcolor{black}\href{https://doi.org/10.1103/PhysRevD.98.065016}{\detokenize{10.1103/PhysRevD.98.065016}}}.

\bibitem[Alpeggiani \em{et~al.}(2018)Alpeggiani, Bliokh, Nori, and
  Kuipers]{ABN18}
Alpeggiani, F.; Bliokh, K.Y.; Nori, F.; Kuipers, L.
\newblock Electromagnetic Helicity in Complex Media.
\newblock {\em Phys. Rev. Lett.} {\bf 2018}, {\em 120},~243605.
\newblock
  doi:{\changeurlcolor{black}\href{https://doi.org/10.1103/PhysRevLett.120.243605}{\detokenize{10.1103/PhysRevLett.120.243605}}}.

\bibitem[V\'azquez-Lozano and Mart\'{\i}nez(2018)]{VEM18}
V\'azquez-Lozano, J.E.; Mart\'{\i}nez, A.
\newblock Optical Chirality in Dispersive and Lossy Media.
\newblock {\em Phys. Rev. Lett.} {\bf 2018}, {\em 121},~043901.
\newblock
  doi:{\changeurlcolor{black}\href{https://doi.org/10.1103/PhysRevLett.121.043901}{\detokenize{10.1103/PhysRevLett.121.043901}}}.

\end{thebibliography}

\end{document}